\documentclass{article}

\usepackage[final, nonatbib]{nips_2017}

\usepackage{graphicx}          % Include figure files
\usepackage{amsmath}    % need for subequations
\usepackage{amssymb}    % for symbols
\usepackage{amsthm}
\usepackage[utf8]{inputenc} % allow utf-8 input
\usepackage[T1]{fontenc}    % use 8-bit T1 fonts
\usepackage{hyperref}       % hyperlinks
\usepackage{url}            % simple URL typesetting
\usepackage{booktabs}       % professional-quality tables
\usepackage{amsfonts}       % blackboard math symbols
\usepackage{nicefrac}       % compact symbols for 1/2, etc.
\usepackage{microtype}      % microtypography
\usepackage{appendix}

\title{Unsupervised learning of dynamical and molecular similarity using variance minimization}

\author{
  Brooke E. Husic and Vijay S. Pande\\
  Department of Chemistry\\
  Stanford University\\
  Stanford, CA 94305\\
  \texttt{\{bhusic, pande\}@stanford.edu}\\
}

\begin{document}

\maketitle

\begin{abstract}
In this report, we present an unsupervised machine learning method for determining groups of molecular systems according to similarity in their dynamics or structures using Ward's minimum variance objective function.
We first apply the minimum variance clustering to a set of simulated tripeptides using the information theoretic Jensen-Shannon divergence between Markovian transition matrices in order to gain insight into how point mutations affect protein dynamics.
Then, we extend the method to partition two chemoinformatic datasets according to structural similarity to motivate a train/validation/test split for supervised learning that avoids overfitting.
\end{abstract}

%%%%%%%%%%%%%%%%%%%%%%%%%%%%%%%%%%%%%%%
\section{Introduction}\label{sec:intro}
%%%%%%%%%%%%%%%%%%%%%%%%%%%%%%%%%%%%%%%

Scientists have sought to understand the dynamical behavior of proteins at atomic resolution since the first molecular dynamics (MD) simulation of the 58-amino acid bovine pancreatic trypsin inhibitor (BPTI) in 1977~\cite{McCammon_Nature77}. In the past 40 years, computational chemists have seen major improvements in molecular dynamics methods~\cite{Adcock_ChemRev06a}, and modern datasets can reach biologically relevant timescales (tens of milliseconds using femtosecond time steps) due to specialized hardware~\cite{Shaw_CommunACM08} and distributed computing platforms such as Folding@home~\cite{Shirts_Science00}, GPUGRID~\cite{Buch_JCIM10}, and Google Exacycle~\cite{Kohlhoff_NatureChem14}.

The enormous size of modern MD datasets requires complementary methods to understand and analyze the data in a statistically rigorous way. Markov state models (MSMs) are a popular framework for this type of analysis that use a master equation to represent the thermodynamics and kinetics of a molecular system~\cite{Bowman_Book14}.
%The MSM is created by decomposing the conformations occupied by the simulated molecule into discrete states and then counting the transitions between those states at a regular time interval. Then, state transitions are counted and used to construct a transition probability matrix, which contains conditional transition probabilities on the off-diagonal terms and can be converted to state free energies on the diagonal. The Markovian assumption underlying the MSM is that the probability of transitioning from one state to another state does not depend on the system's history. Thus, MSMs are a powerful way to integrate many short simulations output from a distributed computing network by threading simulations together based on common states~\cite{Singhal_JChemPhys04}.
% The derivation of a variational principle to select the best state decomposition for a MSM~\cite{Noe_MMS13}, combined with cross-validation~\cite{McGibbon_JChemPhys15a}, recommendations for best modeling practices~\cite{Husic_JChemPhys16}, and open source software~\cite{Scherer_JCTC15, McGibbon_JOSS16, Harrigan_BiophysJ17} has enabled a systematic, objective way to construct MSMs for a dynamical system.
Recent advances in MSM applications include complex, multi-system dynamics such as protein-protein association~\cite{Plattner_NatChem17, Zhou_BiophysJ17} or aggregated datasets containing simulations in multiple force fields~\cite{McKiernan_JChemPhys17, Olsson_PNAS17}. It is thus necessary to develop complementary tools for understanding these types of aggregated datasets and quantifying the dynamical similarity among the different systems.
Minimum variance cluster analysis (MVCA), a recently introduced unsupervised learning method to coarse-grain a single MSM, has the versatility to be used both within a single model (for coarse-graining) and among a set of models for dynamical clustering~\cite{Husic_JCTC17b}.
While the authors focus on the coarse-graining application, they conclude with a motivation of the latter application in which they analyze separate folding simulations of a small protein in nine different protein and water force field combinations.
%The authors demonstrate that these nine datasets can be compared based on dynamical similarity, and find differences in the sensitivity of protein force fields to water force fields.

In this report, we focus on the ability of the unsupervised MVCA algorithm to identify groups of molecular systems. We first provide a theoretical background of the MSM transition matrix in order to motivate the development of MVCA for identifying dynamical groups.
We then demonstrate the method on MSMs of capped amino acids to gain insight into protein dynamics for larger systems.
Finally, we showcase the versatility of MVCA by applying it to a completely different problem:~the selection of training, validation, and test sets for cross-validating a chemoinformatic supervised learning task. 

%%%%%%%%%%%%%%%%%%%%%%%%%%%%%%%%%%%%%%%%%%%%%
\section{Theory background}\label{sec:theory}
%%%%%%%%%%%%%%%%%%%%%%%%%%%%%%%%%%%%%%%%%%%%%

%--------------------------------------------------------------------%
\subsection{The Markovian transition matrix} \label{subsec:framework}
%--------------------------------------------------------------------%

We begin with a continuous-time Markovian process, which means that the probability of transitioning from state $x$ to state $y$ after a time interval of $\tau$ does not depend on any states occupied before the system was in $x$. We assume this process is time-homogeneous, stochastic, and reversible with respect to its stationary distribution, $\mu$. The process is also irreducible, which means that there is a path from each state to every other state given sufficient time, and aperiodic.
If we represent our system by a probability distribution $p_t$ at time $t$, and the transition density kernel for $x \rightarrow y$ as $p(x,y)$, we can write the new probability distribution at time $t+\tau$ as,

\begin{align}
p_{t+\tau}(y)/\mu(y) = \int_{\Omega}dx\,p_t(x)p(x,y) = \mathcal{T}(\tau) \circ p_t(y)/\mu(y).
\end{align}

The continuous transfer operator $\mathcal{T}$, which is characterized by its lag time $\tau$, is compact and self-adjoint with respect to the stationary distribution $\mu$. The transfer operator admits a decomposition into eigenvalues and eigenfunctions,

\begin{align}
\mathcal{T}(\tau) \circ \psi_i =& \lambda_i \psi_i, \label{eqn:trans_prop}
\end{align}

\noindent{}where the eigenvalues $\lambda_i$ are real and indexed in decreasing order. The eigenvalue $\lambda_1=1$ is unique and corresponds to the stationary process. All subsequent eigenvalues are on the interval $|\lambda_{i>1}| < 1$ and correspond to dynamical processes in the time series.

To build a MSM, we decompose the subset of conformation space explored by the MD simulation into discrete, disjoint states. By counting the transitions between these states, we calculate the maximum likelihood estimator of the the transition probability matrix to obtain a discrete approximation to $\mathcal{T}$. This transition matrix is the MSM master equation.

%----------------------------------------------------------------------%
\subsection{Distance between MSM transition matrices} \label{subsec:div}
%----------------------------------------------------------------------%

To cluster MSM transition matrices, we review the theory presented in the original MVCA paper~\cite{Husic_JCTC17b}. For two MSMs with row-stochastic transition probability matrices $P$ and $Q$, the divergence from the $i$th row of $Q$ to the $i$th row of $P$ can be written as the Kullback-Leibler divergence,

\begin{align}
\textup{div}_{KL}(P_i||Q_i) \equiv \sum_{j} P_i(j) \log{\frac{P_i(j)}{Q_i(j)}},
\end{align}

\noindent{}where $P_i$ can be thought of as the ``reference'' distribution and $Q_i$ as a ``test'' distribution~\cite{Bowman_JCTC10}. The information theoretic Jensen-Shannon divergence~\cite{Lin_IEEE91} is a related symmetric formulation that utilizes $M$, the elementwise mean of $P$ and $Q$,

\begin{align}
\text{div}_{JS}(P_i||Q_i) \equiv \frac{1}{2}\text{div}_{KL}(P_i||M_i) + \frac{1}{2}\text{div}_{KL}(Q_i||M_i). \label{eqn:js}
\end{align}

\noindent{}Since it has been shown that the square root of~\eqref{eqn:js} obeys the triangle inequality~\cite{Endres_IEEE03}, we can write the following distance metric,

\begin{align}
\text{div}_{\sqrt{JS}}(P_i||Q_i) \equiv \sqrt{\text{div}_{JS}(P_i||Q_i)}. 
\end{align}

\noindent{}Finally, for a scalar distance between the two transition matrices $P$ and $Q$, each of which contains $i$ rows, we define the sum~\cite{Husic_JCTC17b},

\begin{align}
D_{\sqrt{JS}} \equiv & \sum_i \text{div}_{\sqrt{JS}}(P_i||Q_i). \label{eqn:djs}
\end{align}

%----------------------------------------------------------------------%
\subsection{Hierarchical agglomerative clustering} \label{subsec:agglom}
%----------------------------------------------------------------------%

Hierarchical agglomerative clustering is an unsupervised learning method that is initiated with a set of pairwise distances between data points and iteratively merges the two closest clusters or singletons. Hierarchical agglomerative clustering thus requires a similarity function that quantifies the distance between all data points and an objective function that determines how to use those distances to determine which existing clusters to merge at each agglomerating step.
Common examples of objective functions used for hierarchical agglomerative clustering are single, average, and complete linkages, which define the distance between two clusters as the shortest, average, or greatest distance, respectively, between any point in one cluster and any point in the other cluster. Another objective function used for hierarchical agglomerative clustering is Ward's minimum variance criterion~\cite{Ward_JAmerStatistAssoc63}. The agglomeration is performed by merging the two clusters or singletons such that the resulting increase in intra-cluster variance is minimized. Ward's method is usually implemented according to the following recursive distance update~\cite{Mullner_JStatSoft13},

\begin{align} \label{eqn:ward}
d(u,v) =& \sqrt{\frac{|v|+|s|}{T}d(v,s)^2 + \frac{|v|+|t|}{T}d(v,t)^2 - \frac{|v|}{T}d(s,t)^2},
\end{align}

\noindent{}where the clusters $s$ and $t$ have just been merged to create a new cluster, $u$, and the new distance between $u$ and some other cluster $v$ needs to be updated. $|c|$ represents the number of data points contained in cluster $c$, and $T \equiv |s|+|t|+|v|$. A nonrecursive formula for the distance update when $s$ or $t$ is a singleton has also been derived~\cite{Husic_JCTC17}.
We note that the minimum variance formulation is rigorously defined for Euclidean distances, but that the objective function can be used for any similarity function with the understanding that it no longer corresponds to Euclidean variance, which requires the $l_2$-norm. 
%~\cite{Strauss_PlosOne17}.
%For the remainder of this work, we use the term ``minimum variance'' to refer to the motivation of the objective function

In this paper, we first use hierarchical agglomerative clustering with Ward's minimum variance objective function and the $D_{\sqrt{JS}}$ similarity function~\eqref{eqn:djs} to quantify the similarity between multiple models for related dynamical systems.
We then apply Ward's minimum variance objective function with different similarity functions to the hierarchical clustering of molecular structures and reaction fingerprints in order to motivate a cross-validation scheme for a supervised learning model.
%Although we have derived a distance function that obeys the triangle inequality, this is not necessary for agglomerative clustering.

%\section{Results}\label{sec:results}

\section{Results} \label{sec:results}

\subsection{Clustering dynamically restrained tripeptides} \label{sec:trip}

\begin{figure}[ht!]
\centering
\includegraphics[width=0.9\columnwidth]{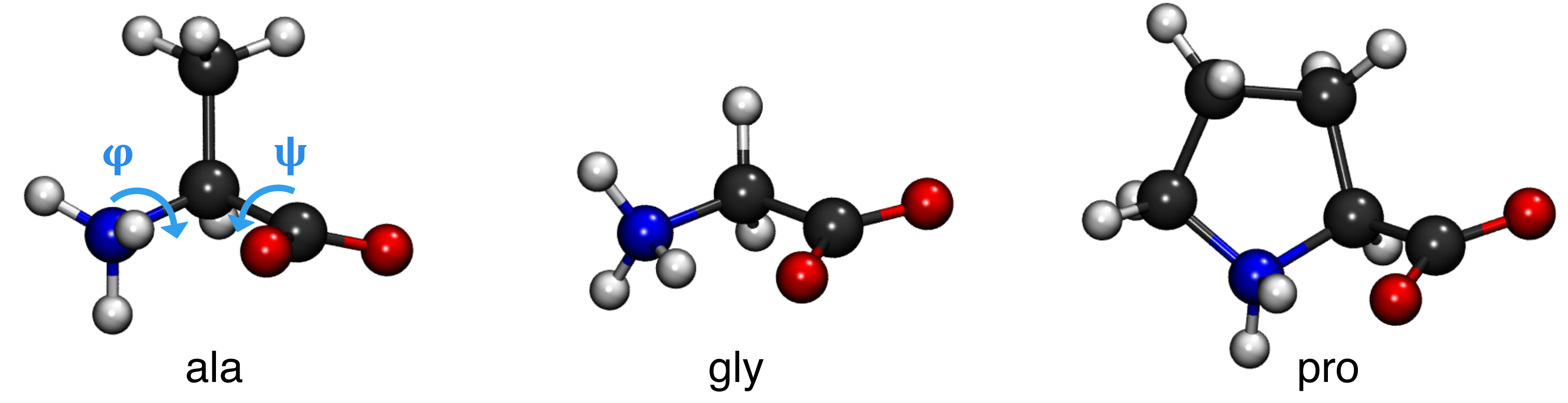}
\caption{Amino acids are differentiated by their side chains, and the dihedrals about the bonds adjacent to the side chains (light blue, left) occupy known regions of $\varphi \times \psi$ space. Alanine (left) is an amino acid with a standard range of side chain motion. Glycine (center) is more flexible because it does not have a side chain. Proline (right) is less flexible because its side chain forms a ring.}
\label{fig:res}
\end{figure}

It is often desirable to modify the dynamics of a protein by introducing a sequence mutation, i.e.~substituting a selected amino acid for the one present in the wild type protein. Simulating mutated proteins using MD can provide a window into whether or not the mutation has affected the protein dynamics. The MVCA algorithm can be used to cluster a set of mutants based on their dynamics, and can lend insight into which mutations produce similar effects. It is known that all solvated amino acids except glycine and proline occupy certain regions of the space defined by their $\varphi$ and $\psi$ backbone dihedral angles~\cite{Vitalini_DataInBrief16}. Figure~\ref{fig:res} shows the $\varphi$ and $\psi$ angles on alanine (left), the structure of which can be compared with glycine (center) and proline (right).
Glycine, the smallest amino acid, occupies more conformations due to its flexibility, and proline occupies fewer conformations due to its 5-member ring. 

\begin{figure}[htb!]
\centering
\includegraphics[width=1.0\columnwidth]{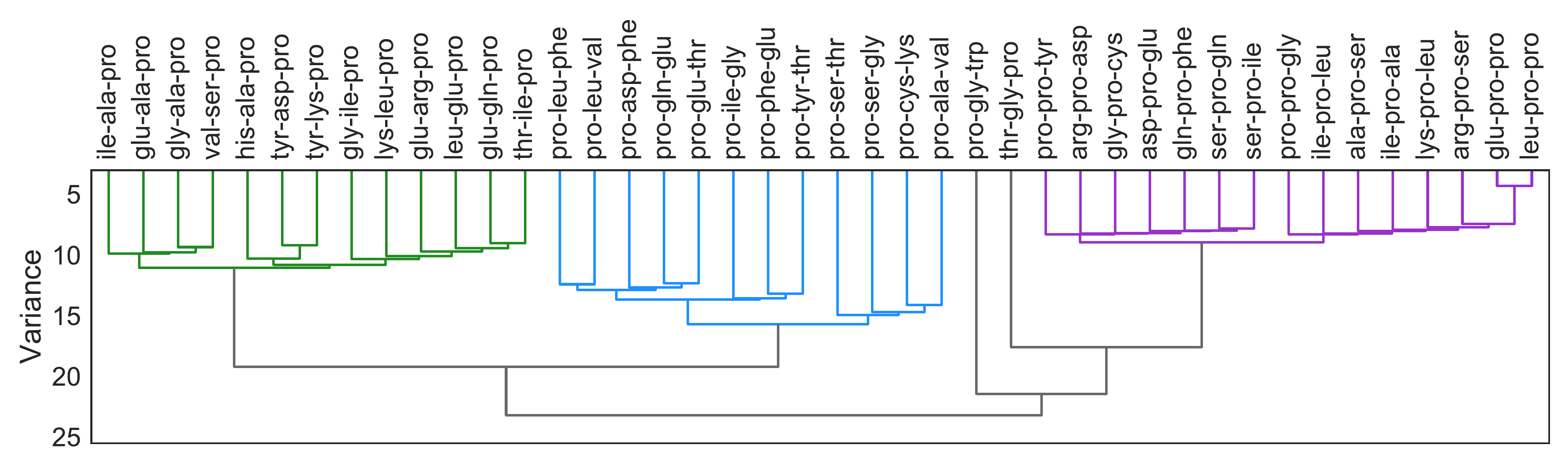}
\includegraphics[width=1.0\columnwidth]{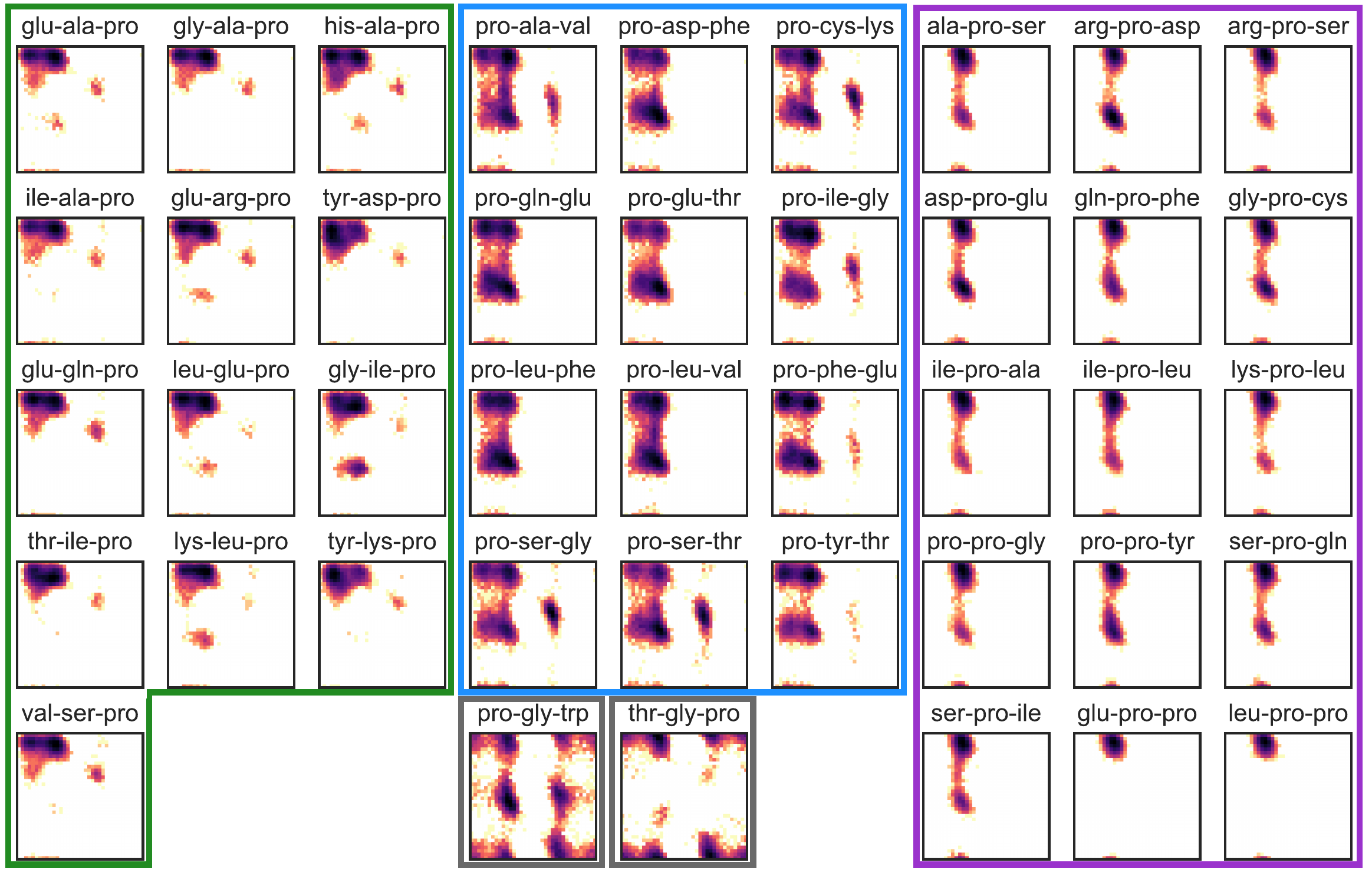}
\caption{A dendrogram from clustering 42 transition matrices using $D_{\sqrt{JS}}$ and Ward's minimum variance objective function shows that the systems cluster into five groups (top). The energy landscape of each system is plotted for $-180 < \varphi < 180$ on the $x$-axis and $-180 < \psi < 180$ on the $y$-axis, with darker colors representing more stable conformations (bottom). The five clusters are identified using boxes corresponding to the colors in the dendrogram. Since we have analyzed a system for which the degrees of freedom are interpretable, we can see that the clustering analysis identifies groups with similar free energy surfaces.}
\label{fig:dend}
\end{figure}

To demonstrate MVCA on a dataset for which we can visualize the dynamical degrees of freedom, we consider a set of $42$ proline-containing tripeptides simulated for 1~$\mu$s each in the AMBER ff99SB-ILDN force field at 300~K.
We hypothesize that the dynamics of the central amino acid in a tripeptide are affected by the presence of proline at any of the three positions.
To investigate the number of dynamical groups represented by this tripeptide dataset, we first create a MSM at a 100~ps lag time for each system using a regular spatial clustering of the $\varphi$ and $\psi$ angles of the central amino acid.
Since all MSMs were built using the same state definitions, we can assess the similarity of their transition matrices using MVCA with $D_{\sqrt{JS}}$.
The MVCA analysis shows that the 42 tripeptides cluster into five natural groups, which is clear from a dendrogram representation of the hierarchical tree (Fig.~\ref{fig:dend}, top). Plotting the free energy surfaces of the central amino acid for each tripeptide shows that the energy landscapes within each cluster are very similar. Three large clusters identify tripeptides with proline as the first, second, and third amino acid (Fig.~\ref{fig:dend}; bottom; blue, purple, and green, respectively). Two singleton clusters are also identified, which represent the two systems in the dataset for which glycine is the central amino acid (Fig.~\ref{fig:dend}, gray). It is clear from their free energy landscapes that these systems are dynamically very different from the others.

We suspect that the systems will differentiate according to the location of the proline because it is easy to understand the degrees of freedom for this simple system. 
This analysis is important because it has fidelity to the expected result, and can successfully group the tripeptides based on only the transition matrices of their MSMs.
The analysis does not produce a reasonable result when the single, average, or complete linkage objective function is used instead of Ward's minimum variance objective function (see Appendix~\ref{appendix}).
%In all three cases, the clustering algorithm cannot separate the system dynamics according to the location of the proline, and can only separate the two glycine-containing tripeptides from the other 40 systems. All three alternative objective functions identify many other singletons, and no coherent groups can be obtained from the results.

\subsection{Clustering small molecules for supervised machine learning}

Predicting properties from molecular structures such as solubility and binding affinity is a significant challenge, and state of the art approaches such as atomic convolutional networks~\cite{Gomes_Arxiv17} and graph convolutions~\cite{Kearnes_JCAMD16} have been used for supervised learning of these quantities.
When using supervised learning to predict molecular properties from a representation of the molecular structure, it is important to use cross-validation when assessing the model's accuracy.
A standard cross-validation split involves dividing the labeled data into three sets: training, validation, and test. The model is trained on the training set and is evaluated on the validation set during development. Once hyperparameters have been selected, the final model is evaluated on the test set.

For chemoinformatic models designed to predict the property of extremely novel new compounds, care must be taken in choosing
% It is not always straightforward to choose
how to divide the data into these three sets such that the model is not overfit to the training data.
When using neural network architectures, for example, the goal is to produce a model that has learned some complex underlying feature from the data, but not a model that has memorized every data point and simply recalls what it has memorized.
A train/validation/test split motivated by differentiating these two options is therefore critical for evaluating supervised learning models in the context of characterizing the properties of novel compounds.
Common partitions for chemoinformatic studies include random splits, temporal splits, stratified splits according to the quantity being learned, and scaffold splits~\cite{Bemis_JMedChem96} in which molecules are assigned to a set based on the frequency of the molecular scaffold. The latter method is generally difficult for models with deep architectures, since the model must apply what it has learned to a different type of data~\cite{Gomes_Arxiv17}.
A ``realistically novel'' training/test split for kinase virtual screening has also been recently published with the goal of avoiding overfit models~\cite{Martin_JCIM17}.

\begin{figure}[htb!]
\centering
\includegraphics[width=1.0\columnwidth]{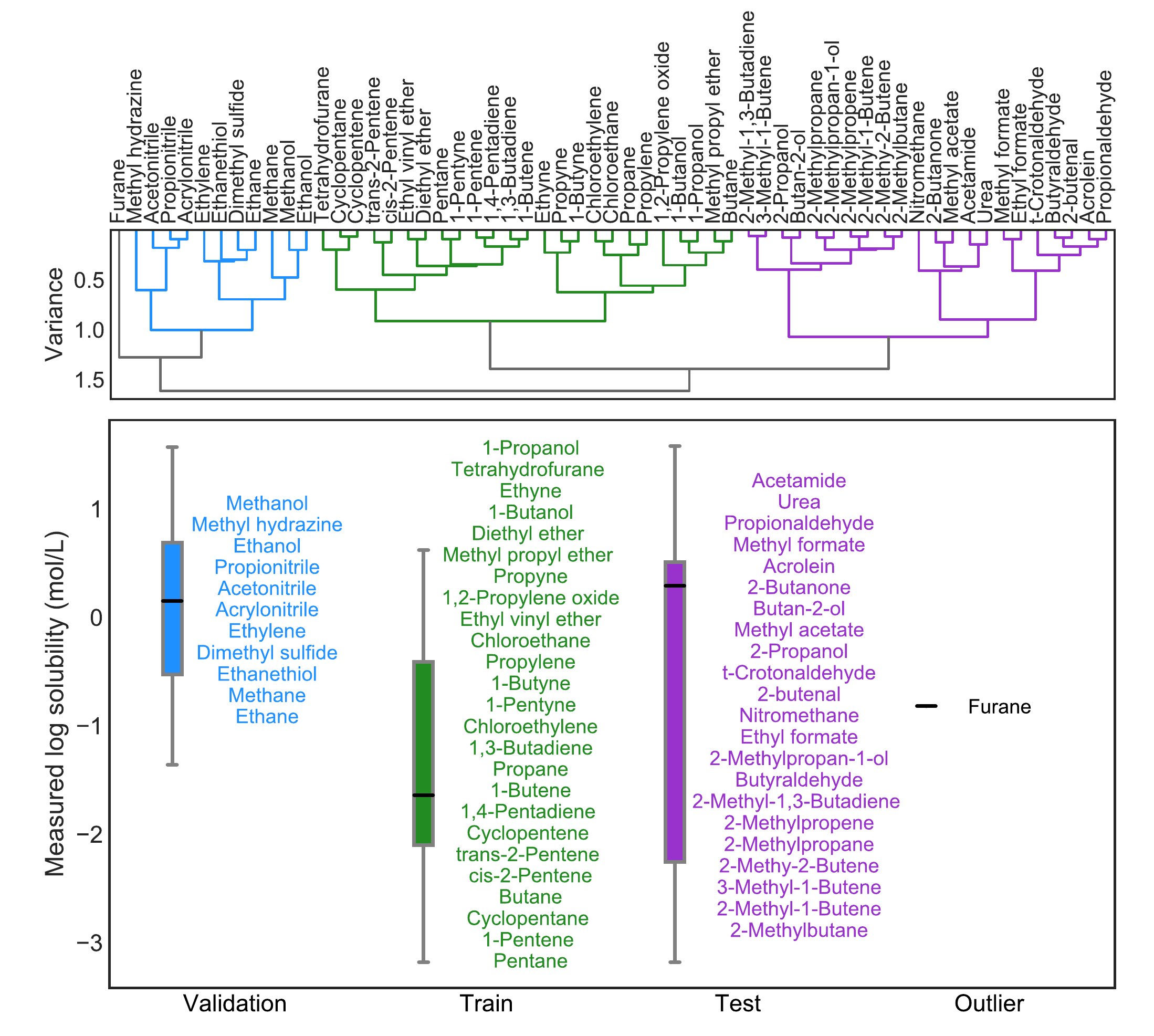}
\caption{A dendrogram from clustering 59 small molecules using Levenshtein distance between their SMILES representations and Ward's minimum variance objective function shows that the molecules cluster into four groups (top). The solubilities of the groups are shown in the box plot (bottom). We identify training (green), validation (blue), and test (purple) sets such that the development set (training and validation) and the test set each span the range of solubilities in the dataset. We note that the structural groups separate by bond saturation, presence of nitrogen or sulfur, and molecular weight, but do not separate alkanes, alkenes, and alcohols. Furane is an outlier based on this analysis.}
\label{fig:box}
\end{figure}

In the spirit of scaffold and ``realistically novel'' splittings, here we apply MVCA to choose the training, validation, and test subsets from a molecular database.
Although it is much too small to use for training a deep network with many hyperparameters, we present an example for a reduced group of molecules so we can interpret the results.
We thus select the 59 compounds with molecular weight less than 75 g/mol from a dataset containing aqueous solubilities~\cite{Delaney_JCICS04}.
In this dataset, the molecules are represented by the simplified molecular-input line-entry system (SMILES), which are one-dimensional string representations of molecular structures.
To quantify the similarity between each pair of molecules, we use the Levenshtein distance ratio between the SMILES strings, which is a function of the number of single character edits that must be made to convert one representation to the other and is normalized for string length.
Since the Levenshtein distance ratio ranges from 0 to 1, where a value of 1 means the strings are identical, we encode the similarity between each pair of strings as the Levenshtein distance ratio subtracted from 1 so that smaller values correspond to closer strings.
Then, we apply MVCA to the pairwise SMILES representations (instead of pairwise $D_{\sqrt{JS}}$ values as in the previous section) and cluster the molecules according to Ward's minimum variance objective function.

Based on the dendrogram representation (Fig.~\ref{fig:box}, top), we then partition the dataset into four groups.  
We see from Fig.~\ref{fig:box} (bottom) that three of the variance-minimized groups according to Levenshtein distance among their SMILES representations contain 11, 25, and 22 molecules, and one group is a singleton containing just furane.
From this partitioning, we would use the largest group as the training set, the 11-member group as the validation set, and the 22-member group as the test set, since the development set (i.e., training and validation) and the test set each span nearly the full range of solubilities. We might choose to include furane in the test set. 
As in Sec.~\ref{sec:trip}, the single, average, and complete linkage functions fail to produce suitable groups for this application (see Appendix~\ref{appendix}).
%While all three objective functions identified furane as a singleton, they displayed the same pathologies noted in the previous section for the tripeptide dataset by isolating many singleton clusters and producing no easily identifiable groups.

\begin{table}[t]
  \caption{Distribution of reactions from 10 classes into two groups determined by MVCA using the Tanimoto coefficient between 4096-bit structural reaction fingerprints. All 10 reaction classes are divided amongst the two groups, and most group allocations are similar to the 78\%/22\% overall splitting.}
  \label{table:rxns}
  \centering
  \begin{tabular}{rlccc}
    \toprule
    Class & Reaction & Quantity & Percent in group 1 & Percent in group 2 \\
     \midrule
    1 & Heteroatom alkylation and arylation & 569 & 79 & 21   \\
    2 & Acylation and related processes & 208 & 62 & 38   \\
    3 & Carbon-carbon bond formation & 133 & 75 & 25   \\
    4 & Heterocycle formation & 31 & 74 & 26   \\
    5 & Protections & 27 & 93 & 7   \\
    6 & Deprotections & 244 & 70 & 30   \\
    7 & Reductions & 487 & 83 & 17   \\
    8 & Oxidations & 86 & 78 & 22   \\
    9 & Functional group interconversion & 208 & 85 & 15   \\
    10 & Functional group addition & 7 & 71 & 29   \\
    \midrule
    & Total & 2000 & 78 & 22 \\
    \bottomrule
  \end{tabular}
\end{table}

What can we learn about molecules represented by SMILES strings from this analysis, and how can we use it to improve the supervised prediction of molecular properties such as solubility for novel compounds?
We see that the training set contains mostly molecules with unsaturated bonds, more than half of the validation set contains compounds with nitrogen or sulfur, and the test set contains larger molecules (no molecular weight lower than 56 g/mol). 
However, we also see that all three non-singleton sets contain alkanes, alkenes, and alcohols.
In terms of training a supervised model, it is important to quantify which SMILES representations are the most similar and can be grouped together with Ward's minimum variance objective function, and using these groups to design training, validation, and test sets will help determine if a model is suitable to predict properties for different kinds of molecular structures or if it is overfitting.
%if a model generalizes well to different kinds of molecular structures or if it is overfitting.
Unlike scaffold splitting, this cross-validation framework is tuned specifically for the choice of molecular representation.

At a higher level, we can use such results to assess the advantages and drawbacks of a given molecular representation by comparing the MVCA grouping to how a domain expert would group a set of molecules.
To illustrate this, we ran a separate analysis of chemical reactions obtained from the United States patent literature and processed as described in~\cite{Liu_Arxiv17} for the prediction of reaction products.
Each chemical reaction is classified into one of ten reaction types~\cite{Schneider_JCIM16}.
We create structural fingerprints for each reaction using the \texttt{CreateStructuralFingerprintForReaction} command in the RDKit Python package~[\url{rdkit.org}].
To quantify the similarity between reaction fingerprints we use the Tanimoto coefficient subtracted from 1 so that smaller values correspond to closer distances, as above.
We then use MVCA to group the 2000 reactions with the lowest total molecular weight.
The resulting dendrogram shows two natural groups with 78\% of the reactions in one group and 22\% of the reactions in the other group.
When we calculate the percent allocation of each structural class in Table~\ref{table:rxns}, we see that unsupervised clustering of reaction fingerprints does not partition the reactions according to their classes.

From the perspective of a modeler seeking to design a train/validation/test split for predicting reaction products, there are two choices for designing a cross-validation scheme.
First, the modeler can determine this split according to the reaction classes in order to represent different types of reactions in the train, validation, and test sets.
However, this may result in an overfit model, since the validation and test sets will contain reactions with representations similar to those in the training set according to their Tanimoto coefficients.
Alternatively, the modeler can determine the split according to the similarity of reaction representations. In this case, the train, validation, and test sets would be chosen according to minimum variance groupings of fingerprints. For the reaction dataset analyzed here, each group would likely contain reactions from all 10 classes.
%Therefore, a cross-validation split based on class labels may result in an overfit model, since the validation and test sets will contain structures similar to those in the training set.
Performing the same analysis with the original reaction SMILES strings and the Levenshtein distance ratio produces nearly identical results, since 98\% of the group assignments are the same as in the fingerprint analysis. This is expected because the fingerprints are generated from the SMILES strings. 
The fingerprints used for this analysis were 4096-bit, and the same analysis was performed for fingerprints with sizes $2^7$ through $2^{12}$. The minimum similarity between any pair of MVCA assignments for fingerprints of different sizes was 93\%. The minimum similarity between SMILES string assignments and fingerprint assignments was 92\% for the 128-bit fingerprints.

Unsupervised clustering of reaction fingerprints shows that the calculated similarity in molecular representations may not align with the similarity assessed by a domain expert, and it is important to consider the similarity of the data according to its representation using an appropriate metric.
Choosing a cross-validation scheme according to human intuition may therefore lead to overfit models when the molecular representations are quantitatively similar but heuristically dissimilar.
Since overfit models have been shown to be a problem in chemoinformatic supervised learning tasks~\cite{Martin_JCIM17}, we anticipate this method will be useful to the chemoinformatic machine learning community.

% We performed an analogous analysis of the same reduced dataset by using extended-connectivity molecular fingerprints~\cite{Rogers_JCIM10} instead of SMILES strings and quantifying their similarity with the Tanimoto coefficient. Performing MVCA on this representation also generates four natural groups with different structural trends than in Fig.~\ref{fig:box}. 

% Thus, MVCA can also be used to evaluate different choices of molecular representations by inspecting what molecular properties separate when the molecules are grouped according to minimum variance in their representation. MVCA can also be applied to learned features during the training process, such as learned graph convolutions. In this case, unsupervised clustering of learned features might facilitate interpretation what information the learned representation is encoding.
% MVCA can thus not only be used to identify a robust cross-validation scheme for training a supervised model, but also to evaluate the information encoded in different representations or learned features.

%Both considerations are important when designing a supervised learning task, and the use of MVCA to identify minimum variance groupings of molecular representations can provide insight into this process.

\section{Discussion}

%The versatility of the MVCA algorithm enables a wide range of applications~\cite{Husic_Draft17}.
% ; for example, it has been previously shown that MVCA can also be applied to identify groups of MSMs for different systems based on the dynamical similarity of those systems
%used for cross-validating the model are designed such that it will be evident whether the model can generalize sufficiently well for chemoinformatic applications.
%generalize to new data points.
%such that the model will only perform well if it generalizes well.
% we can define how well a model performs by how well it generalizes and this is a way of measuring that generalization ability
% Furthermore, unsupervised clustering of molecular representations using MVCA can provide insight into information encoded by the representation, or can enable the interpretation of learned features.
In this report, we present the unsupervised learning of peptide dynamics and of molecular structures using MVCA for hierarchical agglomerative clustering with Ward's minimum variance objective function.
We first demonstrate MVCA with the Jensen-Shannon divergence between MSM transition matrices to identify dynamical groups from a dataset of 42 tripeptides containing proline.
Then, we apply MVCA in a completely new way, using the Levenshtein edit distance ratio between SMILES representations of small molecules, and the Tanimoto coefficient between reaction fingerprints, to partition the datasets for cross-validation.
%according to minimum variance.
This analysis is intended to address a knowledge gap specific to supervised machine learning for chemoinformatic analyses designed to predict the properties of novel molecules:~namely, how to perform cross-validation such that model generalizability is properly measured. 
% This analysis is designed to aid the development of a supervised learning task in which the 
Here, we suggest constructing training, validation, and test sets for model cross-validation according to minimum variance in order to assess model performance with a practical test set; i.e., one that contains newly designed compounds.
%used for cross-validating the model are designed such that it will be evident whether the model can generalize sufficiently well for chemoinformatic applications.
While common machine learning wisdom dictates that the training, validation, and test sets should be drawn from the same distribution, for the prediction of a novel compound's chemical properties it is crucial for us to demonstrate that models can generalize to new kinds of molecules.
If it turns out to be the case that maximally novel test sets break such models, then it is important for the field as a whole to question whether supervised machine learning approaches, in particular those with deep neural network architectures, are appropriate for predicting chemical properties of new compounds.

Machine learning plays a crucial role in both modern MD analyses and the prediction of molecular properties from structure. 
%Unsupervised methods such as MVCA can be used to aid the analysis of a set of mutated systems according to their dynamics. For supervised learning, MVCA can augment model design by identifying training, validation, and test set splits such that variance in the input data structure is minimized, such that a generalizable model can be trained.
We thus anticipate that MVCA will be broadly applicable to various applications in machine learning for molecular data.
The MVCA algorithm is available in the open-source MSMBuilder software package~\cite{Harrigan_BiophysJ17}, which was used to build the MSMs in Sec.~\ref{sec:trip}. The MVCA analyses presented in this report can also be implemented using the SciPy Python package~\cite{Jones_Scipy01}.

\subsubsection*{Acknowledgments}

B.E.H. is grateful to Evan Feinberg, Zhenqin Wu, and Bowen Liu for discussions during the preparation of this manuscript.
The authors thank Matt Harrigan for providing the tripeptide dataset.
We acknowledge the National Institutes of Health under No.~NIH R01-GM62868 for funding.
V.S.P. is a consultant \& SAB member of Schrodinger, LLC and Globavir, sits on the Board of Directors of Apeel Inc, Freenome Inc, Omada Health, Patient Ping, Rigetti Computing, and is a General Partner at Andreessen Horowitz. 
An earlier version of this paper was accepted to the Machine Learning for Molecules and Materials Workshop at NIPS 2017 in Long Beach, CA. The workshop proceedings can be found at \url{http://www.quantum-machine.org/workshops/nips2017/}.

{\small
\bibliographystyle{apalike}
% \bibliography{stanford}

\begin{thebibliography}{}

\bibitem[Adcock and McCammon, 2006]{Adcock_ChemRev06a}
Adcock, S.~A. and McCammon, J.~A. (2006).
\newblock Molecular dynamics survey of methods for simulating the activity of
  proteins.
\newblock {\em Chem. Rev.}, 106(5):1589--1615.

\bibitem[Bemis and Murcko, 1996]{Bemis_JMedChem96}
Bemis, G.~W. and Murcko, M.~A. (1996).
\newblock The properties of known drugs. 1. molecular frameworks.
\newblock {\em J. Med. Chem.}, 39(15):2887--2893.

\bibitem[Bowman et~al., 2010]{Bowman_JCTC10}
Bowman, G.~R., Ensign, D.~L., and Pande, V.~S. (2010).
\newblock Enhanced modeling via network theory: Adaptive sampling of {Markov}
  state models.
\newblock {\em J. Chem. Theory Comput.}, 6(3):787--794.

\bibitem[Bowman et~al., 2014]{Bowman_Book14}
Bowman, G.~R., Pande, V.~S., and No{\'e}, F. (2014).
\newblock An introduction to {Markov} state models and their application to
  long timescale molecular simulation.
\newblock volume 797. Springer.

\bibitem[Buch et~al., 2010]{Buch_JCIM10}
Buch, I., Harvey, M.~J., Giorgino, T., Anderson, D.~P., and Fabritiis, G.~D.
  (2010).
\newblock High-throughput all-atom molecular dynamics simulations using
  distributed computing.
\newblock {\em J. Chem. Inf. Model.}, 50(3):397--403.

\bibitem[Delaney, 2004]{Delaney_JCICS04}
Delaney, J.~S. (2004).
\newblock Esol: Estimating aqueous solubility directly from molecular
  structure.
\newblock {\em J. Chem. Inf. Comput. Sci.}, 44(3):1000--1005.

\bibitem[Endres and Schindelin, 2003]{Endres_IEEE03}
Endres, D.~M. and Schindelin, J.~E. (2003).
\newblock A new metric for probability distributions.
\newblock {\em IEEE Transactions on Information Theory}, 49(7):1858--1860.

\bibitem[Gomes et~al., 2017]{Gomes_Arxiv17}
Gomes, J., Ramsundar, B., Feinberg, E.~N., and Pande, V.~S. (2017).
\newblock Atomic convolutional networks for predicting protein-ligand binding
  affinity.
\newblock {\em arXiv preprint arXiv:1703.10603}.

\bibitem[Harrigan et~al., 2017]{Harrigan_BiophysJ17}
Harrigan, M.~P., Sultan, M.~M., Hern{\'a}ndez, C.~X., Husic, B.~E., Eastman,
  P., Schwantes, C.~R., Beauchamp, K.~A., McGibbon, R.~T., and Pande, V.~S.
  (2017).
\newblock {MSMBuilder:} {Statistical} models for biomolecular dynamics.
\newblock {\em Biophys. J.}, 112(1):10--15.

\bibitem[Husic et~al., 2017]{Husic_JCTC17b}
Husic, B.~E., McKiernan, K.~A., Wayment-Steele, H.~K., Sultan, M.~M., and
  Pande, V.~S. (2017).
\newblock A minimum variance clustering approach produces robust and
  interpretable coarse-grained models.
\newblock {\em J. Chem. Theory Comput.}
\newblock Just Accepted Manuscript.

\bibitem[Husic and Pande, 2017]{Husic_JCTC17}
Husic, B.~E. and Pande, V.~S. (2017).
\newblock Ward clustering improves cross-validated {Markov} state models of
  protein folding.
\newblock {\em J. Chem. Theory Comput.}, 13(3):963--967.

\bibitem[Jones et~al., 2001]{Jones_Scipy01}
Jones, E., Oliphant, T., Peterson, P., et~al. (2001).
\newblock {SciPy}: Open source scientific tools for {Python}.

\bibitem[Kearnes et~al., 2016]{Kearnes_JCAMD16}
Kearnes, S., McCloskey, K., Berndl, M., Pande, V., and Riley, P. (2016).
\newblock Molecular graph convolutions: moving beyond fingerprints.
\newblock {\em J. Comput. Aided Mol. Des.}, 30(8):595--608.

\bibitem[Kohlhoff et~al., 2014]{Kohlhoff_NatureChem14}
Kohlhoff, K.~J., Shukla, D., Lawrenz, M., Bowman, G.~R., Konerding, D.~E.,
  Belov, D., Altman, R.~B., and Pande, V.~S. (2014).
\newblock Cloud-based simulations on {Google} {Exacycle} reveal ligand
  modulation of {GPCR} activation pathways.
\newblock {\em Nature Chem.}, 6(1):15--21.

\bibitem[Lin, 1991]{Lin_IEEE91}
Lin, J. (1991).
\newblock Divergence measures based on the {Shannon} entropy.
\newblock {\em IEEE Transactions on Information Theory}, 37(1):145--151.

\bibitem[Liu et~al., 2017]{Liu_Arxiv17}
Liu, B., Ramsundar, B., Kawthekar, P., Shi, J., Gomes, J., Nguyen, Q.~L., Ho,
  S., Sloane, J., Wender, P., and Pande, V. (2017).
\newblock Retrosynthetic reaction prediction using neural sequence-to-sequence
  models.
\newblock {\em arXiv preprint arXiv:1706.01643}.

\bibitem[Martin et~al., 2017]{Martin_JCIM17}
Martin, E.~J., Polyakov, V.~R., Tian, L., and Perez, R.~C. (2017).
\newblock {Profile-QSAR} 2.0: Kinase virtual screening accuracy comparable to
  four-concentration {IC50s} for realistically novel compounds.
\newblock {\em J. Chem. Inf. Model.}, 57(8):2077--2088.

\bibitem[McCammon et~al., 1977]{McCammon_Nature77}
McCammon, J.~A., Gelin, B.~R., and Karplus, M. (1977).
\newblock Dynamics of folded proteins.
\newblock {\em Nature}, 267(5612):585--590.

\bibitem[McKiernan et~al., 2017]{McKiernan_JChemPhys17}
McKiernan, K.~A., Husic, B.~E., and Pande, V.~S. (2017).
\newblock Modeling the mechanism of cln025 beta-hairpin formation.
\newblock {\em J. Chem. Phys.}, 147(10):104107.

\bibitem[M{\"u}llner, 2013]{Mullner_JStatSoft13}
M{\"u}llner, D. (2013).
\newblock fastcluster: {Fast} hierarchical, agglomerative clustering routines
  for {R} and {Python}.
\newblock {\em J. Stat. Soft.}, 53(1):1--18.

\bibitem[Olsson et~al., 2017]{Olsson_PNAS17}
Olsson, S., Wu, H., Paul, F., Clementi, C., and No{\'e}, F. (2017).
\newblock Combining experimental and simulation data of molecular processes via
  augmented {Markov} models.
\newblock {\em Proc. Natl. Acad. Sci.}, 114(31):8265--8270.

\bibitem[Plattner et~al., 2017]{Plattner_NatChem17}
Plattner, N., Doerr, S., De~Fabritiis, G., and No{\'e}, F. (2017).
\newblock Complete protein--protein association kinetics in atomic detail
  revealed by molecular dynamics simulations and markov modelling.
\newblock {\em Nat. Chem.}, 9(10):1005--1011.

\bibitem[Schneider et~al., 2016]{Schneider_JCIM16}
Schneider, N., Stiefl, N., and Landrum, G.~A. (2016).
\newblock What's what: The (nearly) definitive guide to reaction role
  assignment.
\newblock {\em J. Chem. Inf. Model.}, 56(12):2336--2346.

\bibitem[Shaw et~al., 2008]{Shaw_CommunACM08}
Shaw, D.~E., Deneroff, M.~M., Dror, R.~O., Kuskin, J.~S., Larson, R.~H.,
  Salmon, J.~K., Young, C., Batson, B., Bowers, K.~J., Chao, J.~C., Eastwood,
  M.~P., Gagliardo, J., Grossman, J.~P., Ho, C.~R., Ierardi, D.~J.,
  Kolossv\'{a}ry, I., Klepeis, J.~L., Layman, T., McLeavey, C., Moraes, M.~A.,
  Mueller, R., Priest, E.~C., Shan, Y., Spengler, J., Theobald, M., Towles, B.,
  and Wang, S.~C. (2008).
\newblock Anton, a special-purpose machine for molecular dynamics simulation.
\newblock {\em Commun. ACM}, 51(7):91--97.

\bibitem[Shirts and Pande, 2000]{Shirts_Science00}
Shirts, M. and Pande, V.~S. (2000).
\newblock Screen savers of the world unite!
\newblock {\em Science}, 290(5498):1903--1904.

\bibitem[Vitalini et~al., 2016]{Vitalini_DataInBrief16}
Vitalini, F., No{\'e}, F., and Keller, B. (2016).
\newblock Molecular dynamics simulations data of the twenty encoded amino acids
  in different force fields.
\newblock {\em Data in Brief}, 7:582--590.

\bibitem[Ward, 1963]{Ward_JAmerStatistAssoc63}
Ward, J.~H. (1963).
\newblock Hierarchical grouping to optimize an objective function.
\newblock {\em J. Amer. Statist. Assoc.}, 58(301):236--244.

\bibitem[Zhou et~al., 2017]{Zhou_BiophysJ17}
Zhou, G., Pantelopulos, G.~A., Mukherjee, S., and Voelz, V.~A. (2017).
\newblock Bridging microscopic and macroscopic mechanisms of {p53-MDM2} binding
  with kinetic network models.
\newblock {\em Biophys. J.}, 113(4):785--793.

\end{thebibliography}

}

\appendix

\section{Clustering with other objective functions} \label{appendix}

It is interesting to contrast single, average, and complete linkage functions for hierarchical agglomerative clustering with Ward's method. The following analysis was performed using Scipy~\cite{Jones_Scipy01}.

\begin{figure}[h!]
\centering
\includegraphics[width=1.0\columnwidth]{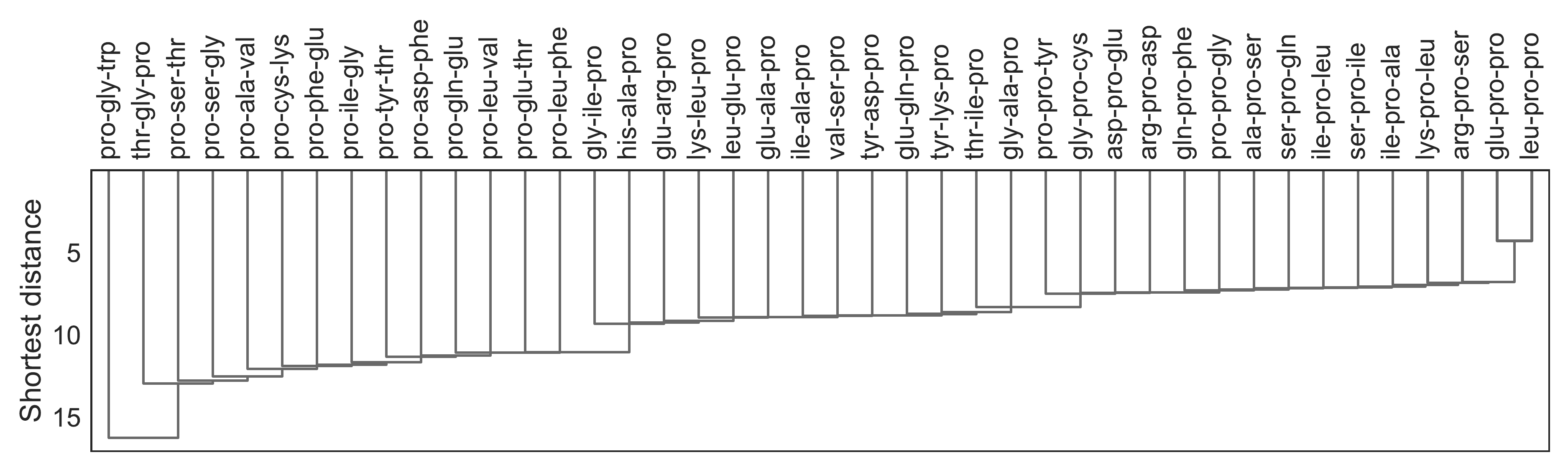}
\includegraphics[width=1.0\columnwidth]{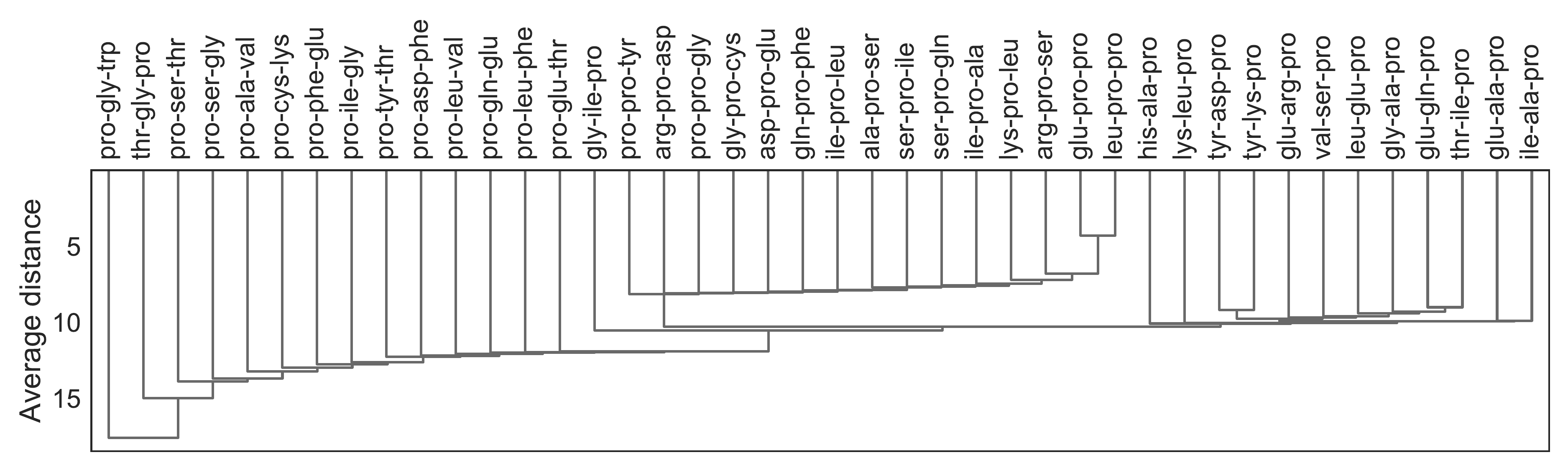}
\includegraphics[width=1.0\columnwidth]{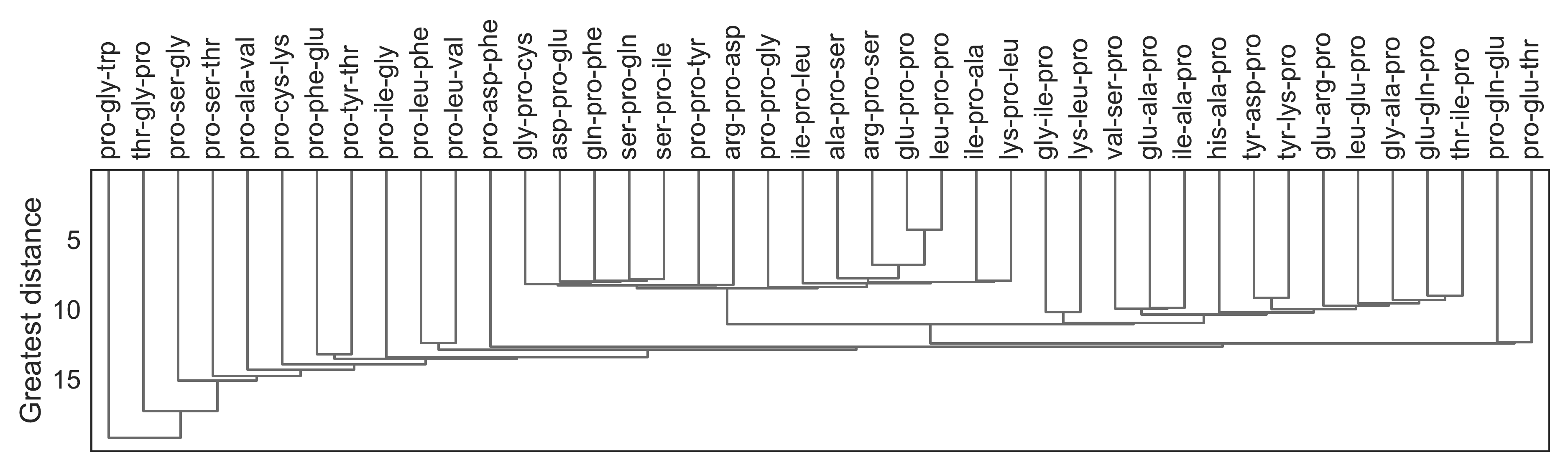}
\caption{Dendrograms created from clustering 42 transition matrices using $D_{\sqrt{JS}}$ and single (top), average (center), and complete (bottom) linkage objective functions. These dendrograms, which do not identify intuitively natural groups, can be contrasted with Fig.~\ref{fig:dend} (top).
In all three cases, the clustering algorithm cannot separate the system dynamics according to the location of the proline, and can only separate the two glycine-containing tripeptides from the other 40 systems. All three alternative objective functions identify many other singletons, and no coherent groups can be obtained from the results.}
\label{fig:a1}
\end{figure}

\begin{figure}[h!]
\centering
\includegraphics[width=1.0\columnwidth]{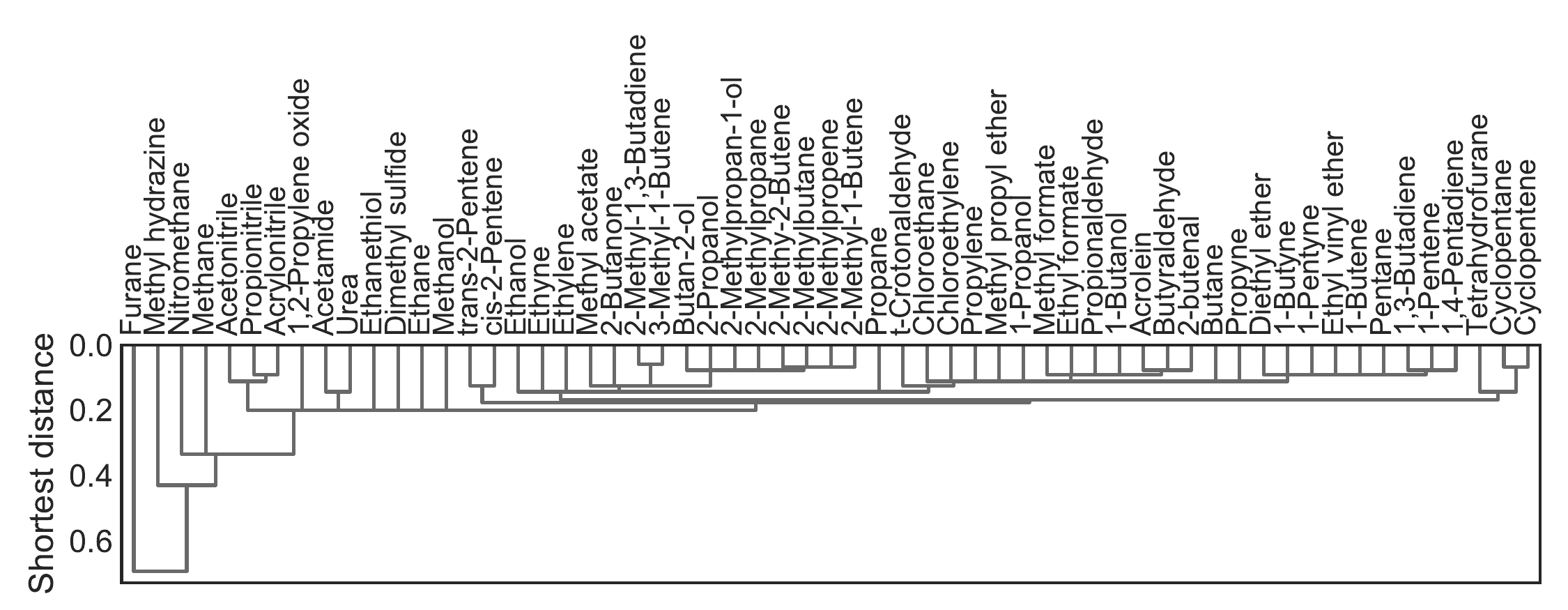}
\includegraphics[width=1.0\columnwidth]{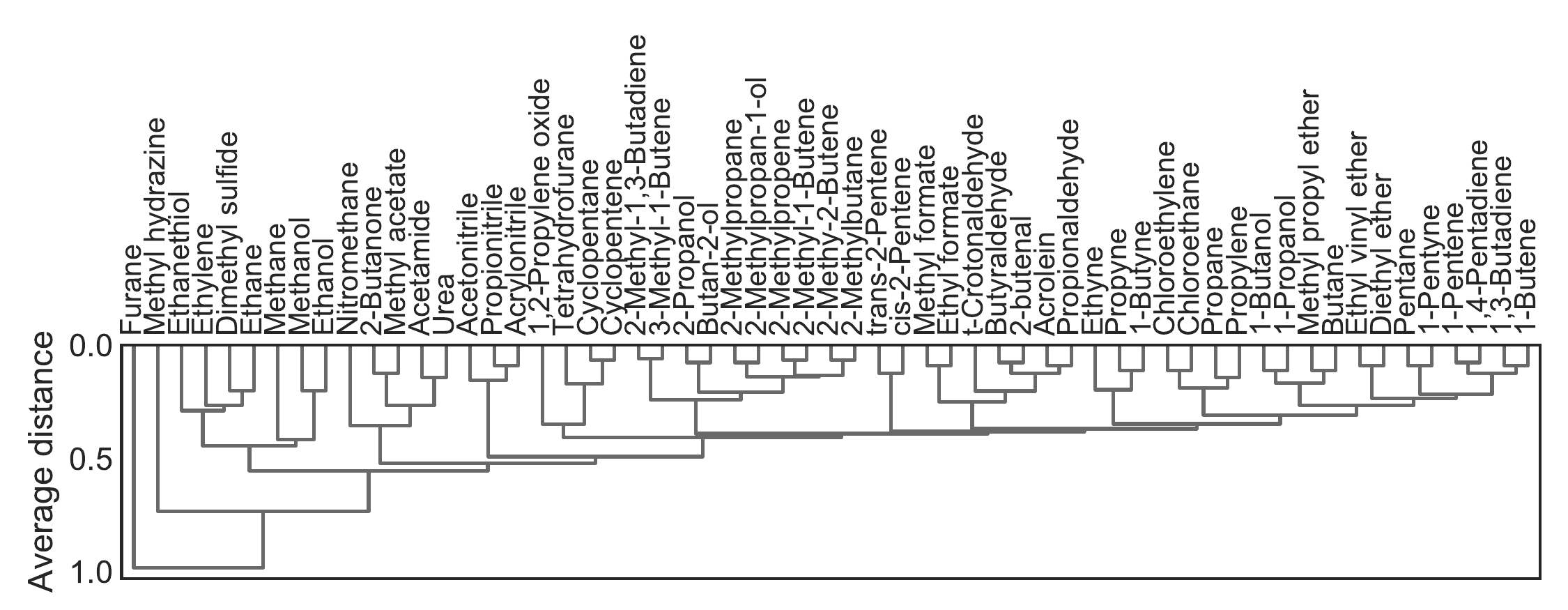}
\includegraphics[width=1.0\columnwidth]{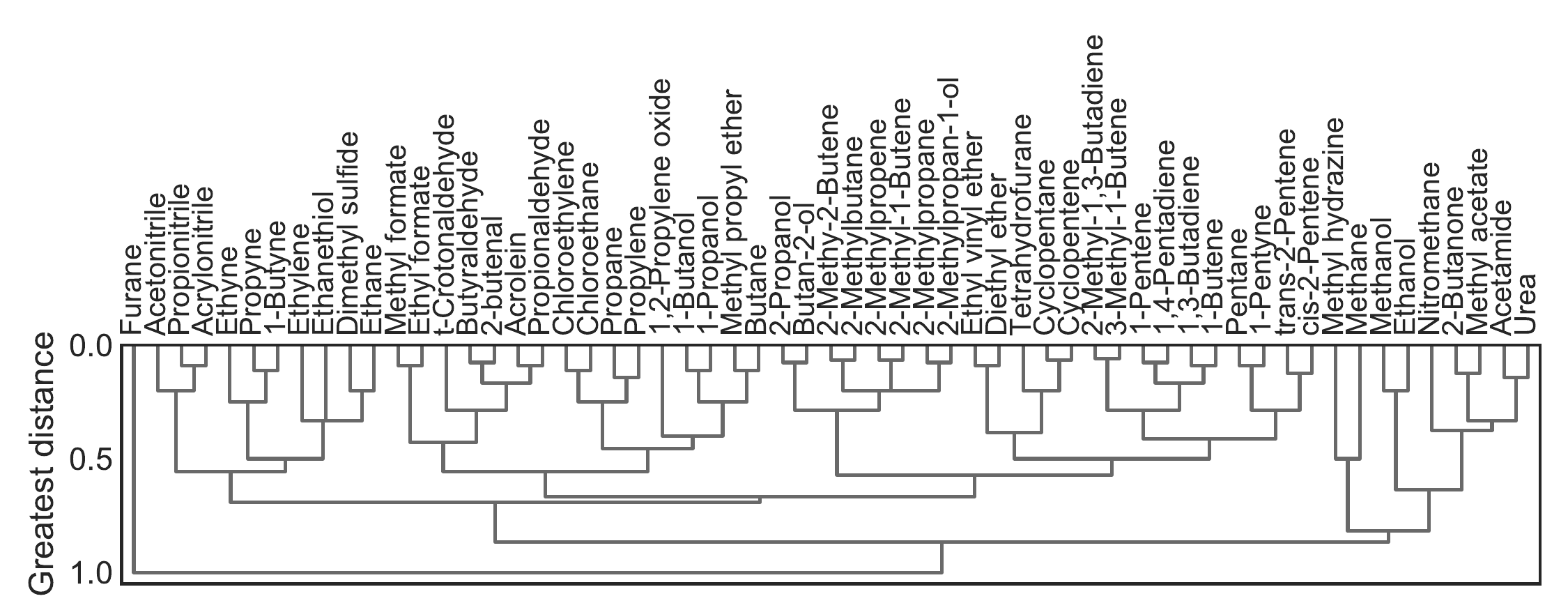}
\caption{Dendrogram created from clustering 59 small molecules using Levenshtein distance between their SMILES representations and single (top), average (center), and complete (bottom) linkage objective functions. These dendrograms, which do not identify intuitively natural groups, can be contrasted with Fig.~\ref{fig:box} (top).
While all three objective functions identified furane as a singleton, they
%displayed the same pathologies noted in the previous section for the tripeptide dataset by isolating many singleton clusters and
produce no easily identifiable groups.}
\label{fig:a2}
\end{figure}

\end{document}